\documentclass{phbauth}
\usepackage{graphicx}

\begin{document}

\begin{frontmatter}

\title{Anisotropic magnetoresistive properties of 
La$_{1-\mathrm{x}}$Ca$_{\mathrm{x}}$MnO$_{3}$ ($\mathrm{x} \approx 1/3$)
film at temperatures far below the Curie temperature}

\author[address1]{B. I. Belevtsev\thanksref{thank1}},
\author[address1]{V. B. Krasovitsky},
\author[address2]{D. G. Naugle},
\author[address2]{K. D. D. Rathnayaka},
\author[address2]{A. Parasiris}

\address[address1]{B. Verkin Institute for Low Temperature Physics \& Engineering, 
Kharkov, 310164, Ukraine}

\address[address2]{Department of Physics, Texas A\&M University, College Station,
TX 77843-4242, USA}

\thanks[thank1]{Corresponding author. E-mail: belevtsev@ilt.kharkov.ua}

\begin{abstract}
A sharp distinction between magnetoresistance (MR) behavior for the 
magnetic fields applied perpendicular ($H_{\bot}$) and parallel 
($H_{\|}$) to the film plane is found in  colossal-magnetoresistance film 
La$_{1-\mathrm{x}}$Ca$_{\mathrm{x}}$MnO$_{3}$ ($\mathrm{x} \approx
1/3$). At increasing of $H_{\bot}$  
the MR is first negative (at $H_{\bot} \leq 4$~kOe), then positive 
(4 kOe $\leq H_{\bot} \leq 12$~kOe), and then negative again
($H_{\bot} > 12$~kOe).  At increasing of $H_{\|}$ the MR is positive below 
$H_{\|} \simeq 6$~kOe and negative above it. In both cases the magnetic field 
was perpendicular to the current. The anisotropic behavior of this kind 
occurs only at low temperatures ($T \leq 18$~K) and is quite different
from the results of previous studies. 
\end{abstract}

\end{frontmatter}

The doped manganites La$_{1-\mathrm{x}}$A$_{\mathrm{x}}$MnO$_{3}$
(where $A$ is a alkaline-earth element) have attracted
great interest recently due to the colossal magnetoresistance (CMR)
observed in these materials at the doping range $0.2 < \mathrm{x} < 0.5$. 
\par 
The bulk manganites have nearly cubic symmetry and therefore should
not have any marked magnetoresistance (MR) anisotropy. In contrast, the CMR films possess
pronounced MR anisotropy in low magnetic fields \cite{eck,wang}.
Resistance $R$ of manganites in ferromagnetic state is a function of
magnetization $\mathbf{M}$ which in turn depends on the temperature
and magnetic field: $R = f[\mathbf{M}(T,\mathbf{H})]$.  Thus the MR
anisotropy in CMR films is in fact some reflection of
$\mathbf{M}(T,\mathbf{H})$ behavior. Two main sources of MR
anisotropy in ferromagnetic films can be pointed out: (1) The
existence of preferential directions of magnetization (due to strains
stemming from the lattice film-substrate mismatch or other
reasons). (2) Dependence of resistance on the angle between current
and magnetization.
\par
It is known\cite{eck,wang} that if easy magnetization axis is
parallel to the film plane, the CMR films show a positive MR when
magnetic field is perpendicular to the film plane, while
for parallel field the MR is negative. This behavior can be associated
with concurrent influence of the above-mentioned anisotropy sources.
In present study, we have found that MR anisotropy in CMR films can
manifest itself in far more complex and puzzling way.


\begin{figure}[t,scale=2]
\begin{center}\leavevmode
\includegraphics[width=0.75\linewidth]{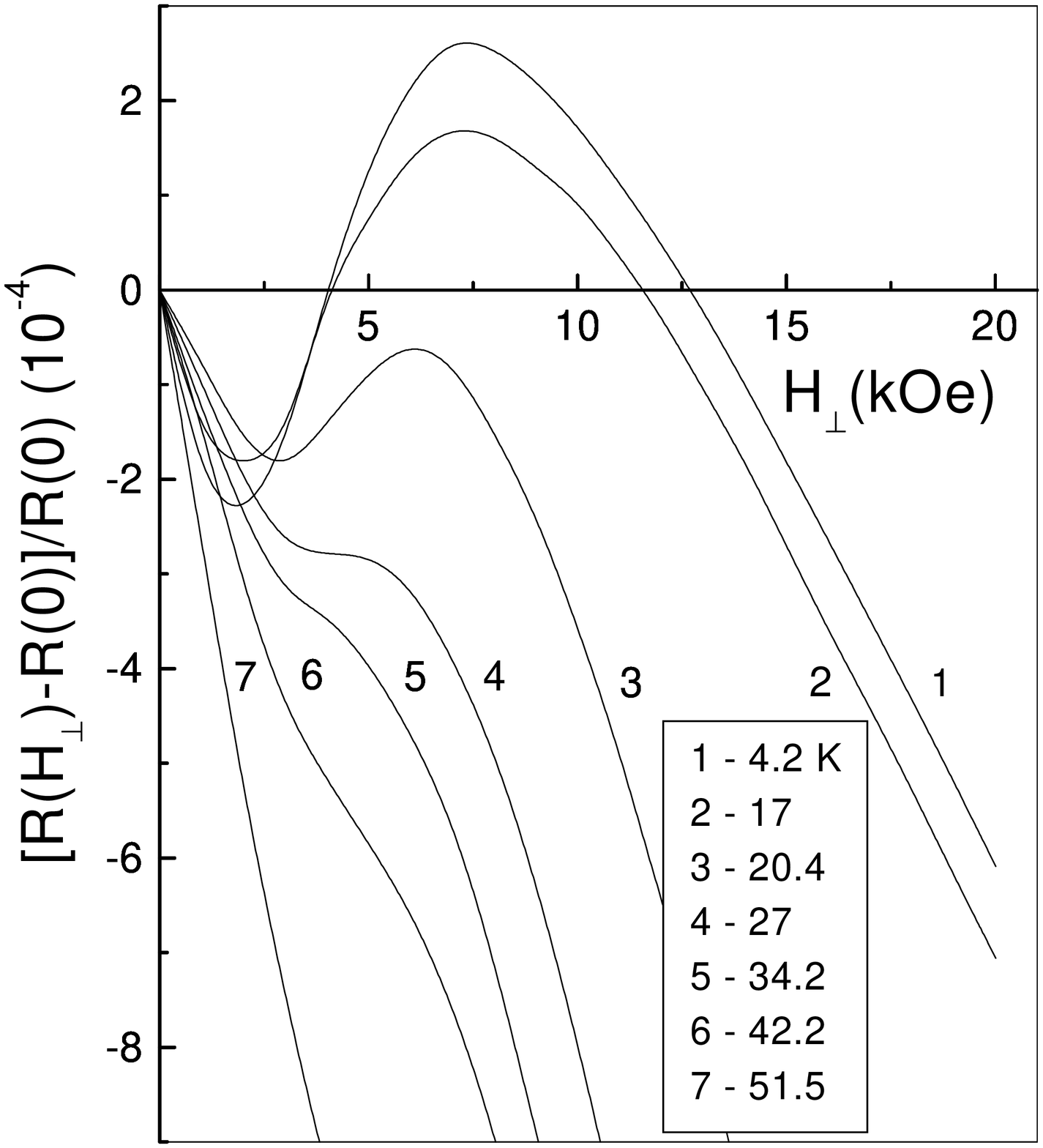}
\caption{ 
Magnetoresistance at different temperatures for the field $H_{\bot}$ 
applied perpendicular to the film plane. 
}\label{fig1}\end{center}\end{figure}
\par 
The object of investigation was
La$_{1-\mathrm{x}}$Ca$_{\mathrm{x}}$MnO$_{3}$ film with $\mathrm{x}
\approx 1/3$ ($\approx 120$~nm thick) which was grown on LaAlO$_{3}$ 
substrate by pulsed-laser deposition. 
The Curie temperature of the film was about 260 K. The MR behavior
for the magnetic fields applied perpendicular ($H_{\bot}$) and
parallel ($H_{\|}$) to the film plane is shown in Figs. 1 and 2 (in
both cases the magnetic field $\mathbf{H}$ was perpendicular to the
current).  At increasing of $H_{\bot}$ the MR is first negative (at
$H_{\bot} \leq 4$~kOe), then positive (4 kOe $\leq H_{\bot} \leq
12$~kOe), and then negative again. At increasing of $H_{\|}$ the MR
is positive below $H_{\|} \simeq 6$~kOe and negative above it. The
anisotropic behavior of this kind occurs only at low temperatures.
At $T > 20$~K the MR is negative for both directions of magnetic
field (Figs.1 and 2). Besides, we found that in the liquid-helium 
temperature range the resistance depends on
angle $\theta$ between $H_{\|}$ and current $I$ approximately in the
following way: $R({\theta})/R(0) = 1 + b\sin^{2}\theta$, where $b >
0$.
\par
The observed puzzling behavior of MR anisotropy is undoubtedly
connected with some interplay of different competing anisotropy
energies in the film studied (shape, strain-induced and intrinsic 
magnetocrystalline anisotropy energies). Besides, the surface and inner
layers of ferromagnetic film may have different anisotropy
properties, that can lead in some cases to canted magnetization and to a 
reorientation of spontaneous magnetization with varying the temperature
or magnetic field\cite{usadel}. The influence of the angle between 
current and magnetization can also give a substantial contribution. 
Taking all these into account and particularly the possible bilayer 
magnetic structure of ferromagnetic film\cite{usadel} it is possible to 
justify the observed results. The precise $\mathbf{M}(T,\mathbf{H})$ 
measurements are needed however to validate any  
particular explanation. We plan to do this.  
\par
We thank S.R. Surthi and R.K. Pandey for film preparation. Support at
TAMU was provided by the Robert A. Welch Foundation (Grant No. A-0514) and the
Texas Advanced Research Program (Grant No. 010366-003).  BIB and DGN acknowledge support
by NATO Scientific Division (Collaborative Research Grant No. 972112).

\begin{figure}[btp]
\begin{center}\leavevmode
\includegraphics[width=0.75\linewidth]{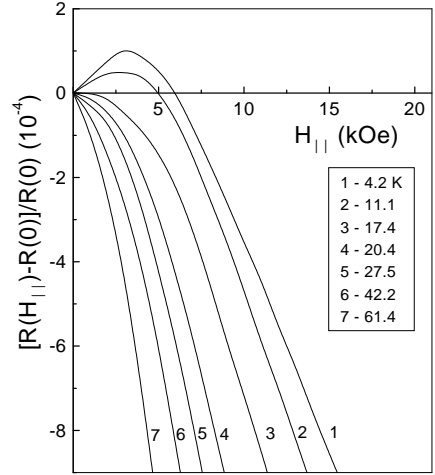}
\caption{ 
Magnetoresistance at different temperatures for the field $H_{\|}$ 
applied parallel to the film plane. 
}\label{fig2}\end{center}\end{figure}

\end{document}